\documentstyle[preprint,aps,epsf]{revtex}
\tighten
\draft
\widetext
\input epsf
\preprint{UPR-992-T}
\bigskip
\bigskip
\begin{document}
\title{\large \bf Cosmological Constraints on Tachyon Matter}
\medskip
\author{
Gary Shiu$^{1}$ and Ira Wasserman$^{2}$  \vspace{0.2cm}}
\address{$^1$ Department of Physics and
Astronomy, University of Pennsylvania, Philadelphia, PA 19104 \\
$^2$ Center for Radiophysics and Space Research, Cornell University, Ithaca, NY 14853}
\date{April 29, 2002}
\bigskip
\medskip
\maketitle

\def\kl#1{\left(#1\right)}
\def\th#1#2{\vartheta\bigl[{\textstyle{  #1 \atop #2}} \bigr] }

\def\kl#1{\left(#1\right)}
\def\th#1#2{\vartheta\bigl[{\textstyle{  #1 \atop #2}} \bigr] }
\newcommand{\drawsquare}[2]{\hbox{%
\rule{#2pt}{#1pt}\hskip-#2pt
\rule{#1pt}{#2pt}\hskip-#1pt
\rule[#1pt]{#1pt}{#2pt}}\rule[#1pt]{#2pt}{#2pt}\hskip-#2pt
\rule{#2pt}{#1pt}}

\newcommand{\fund}{\raisebox{-.5pt}{\drawsquare{6.5}{0.4}}}
\newcommand{\Ysymm}{\raisebox{-.5pt}{\drawsquare{6.5}{0.4}}\hskip-0.4pt%
        \raisebox{-.5pt}{\drawsquare{6.5}{0.4}}}
\newcommand{\Yasymm}{\raisebox{-3.5pt}{\drawsquare{6.5}{0.4}}\hskip-6.9pt%
        \raisebox{3pt}{\drawsquare{6.5}{0.4}}}
\newcommand{\antifund}{\overline{\fund}}
\newcommand{\bYasymm}{\overline{\Yasymm}}
\newcommand{\bYsymm}{\overline{\Ysymm}}


\def\be{\begin{equation}}
\def\ee{\end{equation}}
\def\baray{\begin{eqnarray}}
\def\earay{\end{eqnarray}}
\def\Tt{T^{(t)}}
\def\Tm{T^{(m)}}
\def\Ut{U^{(t)}}
\def\pt{p^{(t)}}
\def\rhot{\rho^{(t)}}
\def\Um{U^{(m)}}
\def\rhom{\rho^{(m)}}
\def\Pm{p^{(m)}}
\def\rhotz{\rhot_0}
\def\ptz{\pt_0}
\def\rhomz{\rhom_0}
\def\pmz{\Pm_0}
\def\Tz{T_0}
\def\Ttz{\dot\Tz}
\def\Ttza{\Ttz}
\def\Tto{\dot T_1}
\def\rhoo{\rhot_1}
\def\po{\pt_1}
\def\Uo{{\bf U}^{(t)}_1}
\def\xvec{{\bf x}}
\def\kvec{{\bf k}}
\def\dotprod{{\mbox{\boldmath $\cdot$}}}
\def\delmu{\nabla_\mu}
\def\dellambda{\nabla_\lambda}
\def\delsigma{\nabla_\sigma}
\def\delnup{\nabla^\nu}
\def\delsup{\nabla^\sigma}
\def\grad{{\mbox{\boldmath $\nabla$}}}
\def\Tr{{\rm Tr}}
\def\phih{\phi_H}
\def\psih{\psi_H}
\def\Temp{{\cal{T}}}
\def\Utmu{U^{(t)\mu}}
\def\Utnu{U^{(t)\nu}}
\def\Ttmunu{T^{(t)\mu\nu}}

\begin{abstract}
We examine whether tachyon matter is a viable candidate for
the cosmological dark matter. First, we demonstrate that in order
for the density of tachyon matter to have an acceptable value
today, the magnitude of the tachyon potential energy at the onset
of rolling must be finely tuned. For a tachyon potential
$V(T)\sim M_{Pl}^4\exp(-T/\tau)$, the tachyon must start rolling
at $T\simeq 60\tau$ in order for the density of tachyon matter
today to satisfy $\Omega_{T,0}\sim 1$, provided that standard big
bang cosmology begins at the same time as the tachyon begins to roll.
In this case, the value of $\Omega_{T,0}$ is exponentially sensitive to $T/\tau$ at
the onset of rolling, so smaller $T/\tau$ is unacceptable,
and larger $T/\tau$ implies a tachyon density that is too small
to have interesting cosmological effects.
If instead the universe undergoes a second inflationary epoch after
the tachyon has already rolled considerably, then the tachyon can
begin with $T$ near zero, but the increase of the scale factor
during inflation must still be finely tuned in order for $\Omega_{T,0}
\sim 1$. Second, we show that tachyon
matter, unlike quintessence, can cluster gravitationally on very
small scales. If the starting value of $T/\tau$ is tuned finely enough that
$\Omega_{T,0}\sim 1$, then tachyon matter clusters more or less
identically to pressureless dust. 
Thus, if the fine-tuning problem can be explained,
tachyon matter is a viable candidate for cosmological dark
matter.
\end{abstract}

\section{Introduction}

In brane cosmology, one often assumes that the
initial condition of the universe is a non-BPS brane
configuration, i.e., a brane configuration that breaks supersymmetry.
For example, in the brane inflationary scenario \cite{braneinf}, the
universe starts out with branes and anti-branes,
as well as branes intersecting at angles (supersymmetry is broken
unless branes are intersecting at
specific angles \cite{angles}). The system is
unstable, as is indicated by the presence of tachyon fields.
One can envision that the universe began with such an
unstable configuration, and evolved to a stable configuration
that corresponds to the Standard Model of particle physics.
Indeed some stable configurations which give rise to features
resembling the Standard Model have been constructed recently \cite{CSU}.

Sen \cite{{sen1},{sen2},{sen3}} has pointed out that rolling tachyons can
contribute a mass density to the universe that resembles classical
dust, and Gibbons \cite{gibbons} has considered the cosmological
background equations that would result if there were a dominant
tachyonic component. However, these papers leave two questions
unanswered that are important for cosmology: (1) Under what conditions
could the tachyon matter density be just comparable to the closure
density of the universe, $\rho_c=3H_0^2/8\pi G$, today? (2) Can tachyon
matter cluster gravitationally?

Any viable candidate for the cosmological dark matter must have
a density $\lesssim\rho_c$ today. Moreover, in order for that
component to be consistent with well-documented successes of
early universe cosmology, such as the synthesis of the light
elements, it must not have become dominant too early in the
history of the universe. We shall see that the density associated
with the rolling tachyon decays {\it slower than} $a^{-3}$ as
the universe expands, where $a(t)$ is the cosmological scale
factor (e.g. \cite{gibbons}, and Eq. [\ref{bckgndt}] below),
at {\it all} times, irrespective of the precise form of $a(t)$.
Thus, if the tachyon matter density is $\lesssim\rho_c$
today, it has only been a dominant constituent of the universe
since a temperature $\sim 10$ eV. This means that the tachyon
density at the onset of rolling must have been very small
compared to the total density of the universe at that time,
provided that standard big bang cosmology began at the same time.
Instead, if there was a second inflationary epoch some time 
after the tachyon began to roll, then the tachyon  density
could have been large -- probably even dominant -- during
the phase before the second inflation. Each scenario 
presents a fine-tuning issue that is quantified in
\S~\ref{bcosmo}.

Since the energy density of tachyon matter decays
like $a^{-3}$ at late times (see \cite{gibbons} and
\S~\ref{bcosmo} below), it is important to ask whether
this component of the universe is capable of gravitational
clustering. For example, quintessence fields with effective
potentials $V(\psi)\propto e^{-\psi/\psi_0}$ can also
have energy density $\propto a^{-3}$ at late times, but
are incapable of clustering on any length scale smaller than
the cosmological particle horizon $H^{-1}$. As a result, the
growth of cosmological density perturbations are slowed in
such models for the dark energy of the universe. Consequently, only
a small fraction of the mass density can be in the form
of quintessence fields with exponential potentials in order
for the observed large scale structure of the universe
to arise from fractional density perturbations $\sim 10^{-5}$,
as are indicated by observations of temperature fluctuations
in the cosmic microwave background radiation. In \S~\ref{perts}
we prove that tachyon matter can cluster under its
own self-gravity, on all but very small length scales.
Thus, if the fine-tuning associated with the smooth tachyon
matter density can be rationalized, rolling tachyons can be
a viable candidate for the cosmological dark matter.
\footnote{We focus here on what happens if the tachyon 
evolves quickly as it rolls toward the zero of its potential,
$\dot T\to 1$ as $V(T)\to 0$, as in \cite{{sen1},{sen2},{gibbons}}.
If the tachyon were to roll slowly ($\dot T\ll 1$)
instead, its cosmology
would resemble quintessence to lowest order in
$\dot T^2$ \cite{pad}. That situation
could present different fine-tuning challenges.}

The calculations presented here are very similar to work
completed recently and independently by Frolov, Kofman
\& Starobinsky \cite{FKS}. One minor difference is that the
computation of tachyon
density fluctuation modes in \S~\ref{perts} below
is done in a different gauge than in Ref. \cite{FKS}. In all
other respects, the two calculations agree. In particular,
we find the same minimum scale for perturbations to be
self-gravitating, corresponding to a comoving wavenumber
$k\simeq Ha/\sqrt{1-\Ttz^2}$, independent of the
form of the tachyon potential, 
where $\Ttz$ is the time derivative
of the tachyon field in the background and $H(t)=\dot a/a$ is the
expansion rate of the universe.
A more important difference,
though, is our discussion in \S~\ref{bcosmo}
of the fine-tuning required in order for the tachyon
dark matter density to be really viable. This fine-tuning
is implicit in the observation by Frolov et al. that linear
theory fails to describe tachyon matter density
perturbations quite early in the history of the universe,
for generic initial conditions. Here, we accept the possibility
that very special initial conditions are needed in order for
tachyon matter to have acceptable cosmological consequences.
A more explicit -- but related -- manifestation of the fine
tuning, which we emphasize in \S~\ref{bcosmo},
is that unless the tachyon field begins rolling at
values in a very small range, the matter density today would
be either far too large or uninterestingly small.

\section{Tachyon Cosmology}

\subsection{Field Equations and Conservation Laws}

Consider tachyon matter with an energy-momentum tensor of the
form \cite{{sen1},{sen2},{gibbons}}
\be
\Tt_{\mu\nu}={V(T)\over\sqrt{1+g^{\alpha\beta}T_{,\alpha}T_{,\beta}}}
\left[-g_{\mu\nu}(1+g^{\alpha\beta}T_{,\alpha}T_{,\beta})+T_{,\mu}
T_{,\nu}\right].
\ee
Here, $V(T)$ is a potential of arbitrary form, which we will only
specify toward the end. We can put this into a more suggestive form
if we define
\be
\Ut_\mu=-{T_{,\mu}\over\sqrt{-g^{\alpha\beta}T_{,\alpha}T_{,\beta}}}
\qquad
\rhot={V(T)\over \sqrt{1+g^{\alpha\beta}T_{,\alpha}T_{,\beta}}}
\qquad
\pt=-V(T)\sqrt{1+g^{\alpha\beta}T_{,\alpha}T_{,\beta}}~;
\label{identify}
\ee
then the energy momentum tensor has a perfect fluid form,
\be
\Tt_{\mu\nu}=\pt g_{\mu\nu}+(\rhot+\pt)\Ut_{\mu}\Ut_{\nu}.
\ee
In terms of these definitions, the equation for the tachyon field
$T$,
\be
{1\over\sqrt{g}}\,{\partial\over\partial x^\mu}
\left({\sqrt{g}V(T)g^{\mu\nu}T_{,\nu}\over
\sqrt{1+g^{\alpha\beta}T_{,\alpha}T_{,\beta}}}\right)
=\delmu\left({V(T)g^{\mu\nu}T_{,\nu}\over
\sqrt{1+g^{\alpha\beta}T_{,\alpha}T_{,\beta}}}\right)
=V^\prime(T)\sqrt{1+g^{\alpha\beta}T_{,\alpha}T_{,\beta}}~,
\nonumber
\ee
where $V^\prime(T)\equiv \partial V(T)/\partial T$,
becomes
\be
\Utmu\delmu\rhot+(\rhot+\pt)\delmu\Utmu=0,
\label{energycons}
\ee
where $\nabla_\mu$ is the covariant derivative.
Conservation of energy and momentum, though, would appear to
imply four equations,
\be
\delmu\,\Ttmunu=0,
\ee
which can be expanded out to
\baray
0&=&\Utnu\left[\Utmu\delmu\rhot+\left(\rhot+\pt\right)\delmu\Utmu\right]
\nonumber\\& &
+\left(g^{\mu\nu}+\Utmu\Utnu\right)\delmu\pt
+\left(\rhot+\pt\right)\Utmu\delmu\Utnu
\nonumber\\
&=&\left(g^{\mu\nu}+\Utmu\Utnu\right)\delmu\pt
+\left(\rhot+\pt\right)\Utmu\delmu\Utnu,
\label{momcons}
\earay
using Eq. (\ref{energycons}) to simplify in the final line.
Eq. (\ref{momcons}) only constitutes three independent equations, since
contracting it with $\Ut_\nu$ yields an identity.
After some algebra, the remaining equation can be written in the form
\be
\Utmu\delmu\Utnu+\left(g^{\mu\nu}+\Utmu\Utnu\right)
\delmu\ln s=0,
\label{navierstokes}
\ee
where $s\equiv\sqrt{-g^{\alpha\beta}T_{,\alpha}T_{,\beta}}$~. Eq. (\ref
{navierstokes}) resembles the Navier-Stokes equation of fluid mechanics,
except that it is worthwhile to recall that it only depends on $\delmu T$,
apart from metric factors. If we substitute for $s$ in Eq. (\ref{navierstokes}),
we find the equation
\be
0=\left(g^{\mu\nu}g^{\lambda\sigma}-g^{\nu\lambda}g^{\mu\sigma}\right)
\delsigma T\delmu(\dellambda T)=\delsigma T\left(\delnup\delsup-\delsup
\delnup\right)T,
\label{Tdoteq}
\ee
which vanishes identically, because the covariant curl of the gradient
of a scalar is zero. Thus, as expected, there is only one nontrivial equation
for the dynamics of the tachyon field, Eq. (\ref{energycons}), even though
$\delmu\,\Ttmunu=0$ appears, at first sight, to impose four different
conservation laws on the single scalar field, $T(x)$.

As always, Einstein's field equations are
\be
R_{\mu\nu}-{1\over 2}g_{\mu\nu}R
=-8\pi G T_{\mu\nu},
\ee
adopting sign conventions in Ref. \cite{Weinberg}. This may be rewritten by using
$R=8\pi GT^\lambda_{\,\lambda}$, to find
\baray
R_{\mu\nu}&=&-8\pi G\left(T_{\mu\nu}-{1\over 2}g_{\mu\nu}T^\lambda_{\,\lambda}
\right)\nonumber\\
&=&-8\pi G\left[{1\over 2}\left(\rho-p\right)g_{\mu\nu}+
(\rhot+\pt)\Ut_\mu \Ut_\nu+(\rhom+\Pm)\Um_{\mu}\Um_{\nu}\right],
\earay
where we have also used $T^\lambda_{\,\lambda}=-\rho+3p$, and we include both
tachyonic matter and ordinary matter, whose fluid variables are distinguished
by a superscript $m$. (See \cite{Weinberg},
Eq. [15.1.14] and the equation following his Eq. [15.1.17].)
With the identifications in Eq. (\ref{identify}), we can map conventional
results for the cosmology of perfect fluids to the cosmology of tachyon
matter.

\subsection{Background Cosmology}
\label{bcosmo}

Consider first the background cosmology; for homogeneous, isotropic
expansion, $\Um_\mu=\Ut_{\mu}=(-1,0,0,0)$, and denoting background matter
densities and pressures with subscripts $0$ we obtain the field equations
\baray
H^2&=&{8\pi G(\rhomz+\rhotz)\over 3}\nonumber\\
\dot\rhomz&=&-3H\left(\rhomz+\pmz\right)\nonumber\\
\dot\rhotz&=&-3H\left(\rhotz+\ptz\right),
\label{bckgnd}
\earay
where we have adopted the spatially flat FRW metric
\be
ds^2=-dt^2+a^2(t)\,\delta_{ij}dx^i\,dx^j,
\ee
and dots denote differentiation
with respect to time; as usual, $H\equiv\dot a/a$.
In the cosmological background, the tachyon field is $T_0(t)$, and we
have
\be
\rhotz={V(\Tz)\over\sqrt{1-\Ttz^2}}\qquad
\ptz=-V(\Tz)\sqrt{1-\Ttz^2}
\qquad
\rhotz+\ptz=\Ttz^2\rhotz.
\label{tachprops}
\ee
As has been noted in Refs. \cite{{sen1},{sen2},{gibbons}}, as the tachyon field evolves, it moves
toward $V(\Tz)\to 0$, but retains finite energy density, $\rhotz$, so that
$\Ttz^2\to 1$; thus, at late times, $\dot\rhotz\to -3H\rhotz$, so
$\rhotz\,a^3\to$ constant.\cite{gibbons} Thus, insofar as the background solution is
concerned, the tachyon behaves like a nonrelativistic gas of dust asymptotically.
In this sense, the tachyon field would appear to be a good candidate for the
dark matter.

However, there are constraints on the tachyon field, namely that the density
today must be a fraction $\Omega_{T,0}\leq 1$ of the total mass density of
the universe, and also that the tachyon density was subdominant at early times,
particularly during the nucleosynthesis epoch. To examine these constraints
roughly, suppose that $\Ttz(t)=1-\epsilon(t)$ at all times since well before
nucleosynthesis (perhaps since inflation, for example). Then the density of
tachyons at time $t$ is $\rhotz\simeq V(T_0(t))/\sqrt{2\epsilon(t)}$, and
$T_0(t)\simeq t$. To examine the constraint imposed by the present day mass
density, suppose that tachyon matter comprises a fraction $\Omega_{T,0}$
of the total mass density of the universe today, and that the universe is
flat (as indicated by CMB observations), so the total mass density is
$3H_0^2M_{Pl}^2/8\pi$, where $H_0$ is the Hubble constant, and $M_{Pl}$
the Planck mass. Then the value of $\epsilon$ today must be
\be
\epsilon_{today}\simeq{32\pi^2V_{today}^2\over 9\Omega_{T,0}^2M_{Pl}^4
H_0^4}~;
\ee
clearly, we want $\epsilon_{today}\ll 1$, so we must have
$V_{today}\ll 3\Omega_{T,0}H_0^2M_{Pl}^2/4\pi\sqrt{2}$. Assuming that a
natural scale for $V(T)$ is $M_{Pl}$, we can recast this bound as
$V_{today}/M_{Pl}^4\ll 3\Omega_{T,0}H_0^2/4\pi\sqrt{2}\,M_{Pl}^2
\simeq 2.5\times 10^{-123}\Omega_{T,0}$, since $H_0/M_{Pl}\simeq 1.2\times 10^{-61}$
(i.e. $H_0\simeq 70\,{\rm km~s^{-1}~Mpc^{-1}}$ in Planck units).
Suppose, as a specific example, that $V(T)=M^4\exp(-T/\tau)$, where
$M\sim M_{Pl}$ and
$\tau$ is a characteristic timescale, which might also be expected to be
the Planck scale. Since $T_0(t)\simeq t$, the inequality can be satisfied
as long as the universe is older than about $280\tau$ -- which is true with
plenty of room to spare for any reasonable $\tau$ based on
mass scales of elementary particle physics.

Although it is easy to arrange for $\epsilon_{today}\ll 1$,
considerable fine-tuning is
necessary to achieve $\Omega_{T,0}\sim 1$. Suppose that $\epsilon\sim 1$
at some time in the past, when the temperature of the universe was $T_{start}$
and the tachyon potential was $V_{start}$. Then the tachyon matter density
at this time, which represents the onset of the rolling of the tachyon field,
is $\rhot_{start}\sim 
V_{start}\sim\rho_{today}\Temp_{start}^3/\Temp_{today}^3$, where
$\Temp_{today} \sim 2.73$ K is the
temperature of the universe today,\footnote{$\Temp_{today}$ should be slightly
smaller than the temperature of the universe today ($2.73$ K) 
because of particle annihilations.} provided that the big bang
phase of cosmological evolution begins at $\Temp_{start}$. 
\footnote{This estimate presumes that $\rhot a^3$ is constant. Actually,
for a short period at the start of rolling, $\rhot$ decreases considerably
more slowly than $a^{-3}$. Thus, the estimate of $\rhot_{start}$ we are
using is an overestimate, and the fine tuning problem is actually somewhat
more severe. We note, though, that particle annihilations after 
$\Temp_{start}$ increase the value of $a\Temp$, and this would alleviate
the fine tuning, but only by a factor of at most of order 100 in the value
of $V_{start}$.}
As a result,
\be
{V_{start}\over \Temp_{start}^4}\sim{\rho_{today}\over
\Temp_{today}^3\Temp_{start}}\sim
{100\,{\rm eV}\,\Omega_{T,0}\over \Temp_{start}},
\label{finetune}
\ee
which implies that $V_{start}/\Temp_{start}^4\ll 1$ in order for $\Temp_{start}$
to exceed $\sim 0.1-1$ MeV, the temperature at the nucleosynthesis epoch. This
means that if $\Temp_{start}\sim M_{Pl}$, for example, then $V_{start}/M_{Pl}^4\sim
10^{-26}$.
The requisite tuning is less severe if the fundamental
scale of the theory is smaller, but even for $\Temp_{start}\sim $ TeV, we
must require that $V_{start}/\Temp_{start}^4\sim 10^{-10}$ or so. 
Larger values of $V_{start}/\Temp_{start}^4$ would imply larger values
of $\rho_{today}/\Temp_{today}^3$ than in the observed universe. Smaller
values are consistent with all observed properties of the universe, but
imply very small $\Omega_{T,0}$, so that tachyon matter is not important.

Small values of $V_{start}/\Temp_{start}^4$ can arise if the tachyon begins to roll
from a relatively large value of $T/\tau$.\cite{sen4} For example, if
$V(T)\sim M_{Pl}^4\exp(-T/\tau)$, then the tachyon field must begin to roll
at $T/\tau\simeq 60$ if $\Omega_{T,0}\sim 1$; for a fundamental mass scale
$\sim$ TeV, the roll begins at $T/\tau\simeq 23$. However, if we
adopt $V_{start}/\Temp_{start}^4\sim\exp(-T_{start}/\tau)$, then Eq.
(\ref{finetune}) implies that\footnote{More generally, if 
$V_{start}/\Temp_{start}^4=v(T_{start}/\tau)$, we would find
$$\Omega_{T,0}\sim {\Temp_{start}
v\left({T_{start}/\tau}\right)\over 100\,{\rm eV}}$$
and assuming that $v(x)\to 1$ for $x\ll 1$, and $v(x)\to 0$ for 
$x\to\infty$, then we still must fine tune $T_{start}/\tau$ to a 
large value in order to get $\Omega_{T,0}<1$. If $v(x)$ declines slower
than $e^{-x}$, the values of $T_{start}/\tau$ for which $\Omega_{T,0}\sim
1$ are larger, but the value of $\Omega_{T,0}$ is not as sensitive
to $T_{start}/\tau$.}
\be
\Omega_{T,0}\sim {\Temp_{start}e^{-T_{start}/\tau}\over 100\,{\rm eV}}~.
\ee
Thus, the value of $\Omega_{T,0}$ is exponentially sensitive to
the value of $T_{start}/\tau$, given the values of $\Temp_{start}$
and the known value of $\Temp_{today}$: only a {\it very restricted}
range of $T_{start}/\tau$ is consistent with $\Omega_{T,0}\sim 1$.
A smaller value of $T_{start}/\tau$ would imply $\Omega_{T,0}>1$,
which would be inconsistent with the observed flatness of the
universe (or else inconsistent with the observed $\Temp_{today}$
if we insist on $\Omega_{T,0}\sim 1$, which would be unreasonable).
A larger value of $T_{start}/\tau$ would imply a very small value
of $\Omega_{T,0}$, which would not pose any inconsistencies, but
would be uninteresting, since it would mean that tachyon
matter is only an inconsequential component of the universe.

We have presented this fine tuning from the point of view of 
integrating backward from the present day to the time when the tachyon
starts to roll toward $V=0$. We can also state the same result from
the point of view of integrating forward. Suppose that the tachyon
starts rolling at $\Temp_{start}$, when the potential is $V_{start}
=\lambda\Temp_{start}^4$; generalize the above discussion by also
assuming that $\sqrt{1-\Ttz^2}=2\epsilon_{start}$ at this time.
Then the tachyon matter density is $\rho_{start}=\lambda\Temp_{start}^4/
2\epsilon_{start}$ when the tachyon begins to roll. As long as $\Ttz$
is not too close to zero initially, a rough estimate of the present
tachyon matter density is $\rho_{today}\sim\rho_{start}\Temp_{today}^3/
\Temp_{start}^3$. Thus, we find that $\rho_{today}\sim \lambda \Temp_{start}
\Temp_{today}^3/2\epsilon_{start}$, and since $\rho_{today}/\Temp_{today}^3\sim 100\Omega
_{T,0}$ eV, we conclude that $\lambda/2\epsilon_{start}\sim 100\Omega_{T,0}
{\rm eV}/\Temp_{start}$. For any reasonable value of $\Temp_{start}$, we see
that $\lambda/2\epsilon_{start}\ll 1$ is required in order for 
$\Omega_{T,0}$ to be of order one.

The fine-tuning problem derived above holds if standard big bang
cosmology begins at the same time as the tachyon begins its roll
toward $V=0$. Alternatively, we might imagine that the universe 
undergoes a second inflationary period of exponential expansion
after the tachyon begins to roll. In that case, the universe
could begin with $V_{start}\sim\Temp_{start}^4$, and would become
tachyon dominated soon afterward. The density of tachyonic matter
today would be $\rho_{today}\sim\Temp_{start}^4a_{start}^3$, but
$a_{start}\ll \Temp_{today}/\Temp_{start}$ as a result
of the intervening inflation.  In order to get a particular value
of $\Omega_{T,0}$, we must have
$\Temp_{start}a_{start}^{3/4}\sim 0.003\Omega_{T,0}^{1/4}$ eV;
for $\Temp_{start}\sim M_{Pl}$, we find $a_{start}\sim 10^{-41}
\Omega_{T,0}^{1/3}$, which is $\sim 10^{-9}\Omega_{T,0}^{1/3}
\Temp_{today}/M_{Pl}$.
Although such a model allows $V_{start}\sim\Temp_{start}^4$,
it still presents a fine-tuning problem, since the value of
$\Temp_{start}a_{start}^{3/4}$ must be adjusted precisely to yield
$\Omega_{T,0}\sim 1$. This means that the expansion factor during
the second inflation must be just right; for $\Temp_{start}\sim
M_{Pl}$, the expansion factor must be $\sim 10^9\Omega_{T,0}^{-1/3}$.
Too much second inflation would dilute the tachyon density to
an uninteresting level, whereas too little second inflation would
leave an unacceptably high tachyon density.\footnote{
We also note that density fluctuations in the
tachyon matter could grow during the pre-inflation epoch in
such a model (see \cite{FKS}, and \S~\ref{perts} below).}

This fine-tuning is an unavoidable property of the tachyon dark matter
model. The tachyon matter density $\rhotz$
evolves according to $\dot\rhotz=-3H\Ttz^2\rhotz$ (see Eqs. [\ref{bckgnd}]
and [\ref{tachprops}], as well as Eq. [\ref
{bckgndt}] below; see \cite{gibbons}). Since $\Ttz\leq 1$, we see that $\rhotz$ decreases
{\it slower than} $a^{-3}$ at all times as the universe expands,
irrespective of the time dependence of the cosmological scale factor
$a(t)$. Thus, once the standard big bang phase of early universe
cosmology starts, it is unacceptable for $\rhotz$ to be comparable
to the radiation density in the universe -- if it were, then tachyons
would dominate right away, and there would never be a radiation
dominated phase of expansion. (Subsequent particle annihilations
bump up the density of radiation by numerical factors, but unless
there are enough annihilations, and they are timed propitiously,
premature tachyon domination is still inevitable.) If the 
standard big bang phase of the universe begins at the same time
as tachyons start to roll toward $V(T)\to 0$, then in order
that tachyons only become dominant at a temperature $\sim 10$ eV
the starting value of $V(T)$ must be very small compared to
its ``natural scale'' (perhaps $\sim M_{Pl}^4$). Whatever the
tachyon potential, this requires that $T$ started to roll from
a relatively large value.
Even if there was a subsequent, second epoch of inflation, before
which tachyons were dominant, it must lower the tachyon density enough
that they do not dominate again before a temperature $\sim 10$ eV.
Either scenario presents a fine tuning issue irrespective of the
precise form of the tachyon potential, $V(T)$, if tachyons are
the cosmological dark matter. Tachyon matter could be present
at a negligible density today, in which case the dark matter must
be something else, if either the tachyon began rolling
from a large enough value of $T$, or the tachyon matter density
inflated to a minute level subsequent to the onset of rolling.

\subsection{Do Tachyons Cluster Gravitationally?}
\label{perts}

It is well known, from studies of cosmological perturbation theory,
that a non-particle species of dark energy with $\rho_0 a^3$ is only
an acceptable component of the universe provided that it can cluster
gravitationally. Otherwise, the growth rate of cosmological density perturbations
is truncated: instead of growth proportional to the scale factor $a(t)$, one
find growth proportional to $[a(t)]^\sigma$, with
$\sigma={1\over 4}\left(\sqrt{1+24f_c}-1\right)$, where $f_c$ is the fraction
of the density of the universe (all of which is presumed to diminish as
$a^{-3}$) that can cluster gravitationally. Thus, the key question is
whether tachyon matter clusters gravitationally.

To see whether tachyon matter can cluster gravitationally, let us
anticipate that it does, and consider only a universe that is actually dominated
by it. (We can insert a cosmological constant, or quintessence field, and
ordinary matter later -- the issue is whether or not the tachyon field
clusters under gravity, irrespective of whether the gravity is due to the
tachyonic matter itself, or to other types of matter.) In this case the
background equations Eq. (\ref{bckgnd}) simplify to \cite{gibbons}
\baray
H^2&=&{8\pi G\rhotz\over 3}={8\pi GV(\Tz)\over 3\sqrt{1-\Ttz^2}}\nonumber\\
\dot\rhotz&=&{d\over dt}\left({V(\Tz)\over\sqrt{1-\Ttz^2}}\right)
=-3H\left(\rhotz+\ptz\right)=-{3H\Ttz^2V(\Tz)\over\sqrt{1-\Ttz^2}}
=-3H\Ttz^2\rhotz~.
\label{bckgndt}
\earay
We can find the perturbation equations easily using results given in
\S~ 15.10 in \cite{Weinberg}; these results are valid in synchronous gauge,
and assume, as in Weinberg \cite{Weinberg}, a metric of the form
\be
ds^2=-dt^2+\left[a^2(t)\delta_{ij}+h^{(W)}_{ij}(\xvec,t)\right]dx^i\,dx^j
=-dt^2+a^2(t)\left[\delta_{ij}+h_{ij}(\xvec,t)\right]dx^i\,dx^j,
\ee
where the superscript $W$ denotes Weinberg's definition of the perturbed
metric, which we alter slightly. It doesn't matter that Weinberg's equations
are not written in terms of gauge invariant forms; we can take gauge invariant
combinations later. The proof that tachyon matter can
cluster should not depend on whether we use gauge independent or gauge
dependent variables, although one has to be careful about modes that
appear to grow, but are merely gauge artifacts.

To solve the problem, we introduce a perturbation $T_1$ to the tachyon field,
relative to its background value; associated with it are perturbations
\baray
\rhoo&=&{V^\prime(\Tz)T_1\over\sqrt{1-\Ttz^2}}+{V(\Tz)\Ttz\Tto\over
(1-\Ttz^2)^{3/2}}\nonumber\\
\po&=&-{V^\prime(\Tz)T_1\sqrt{1-\Ttz^2}}
+{V(\Tz)\Ttz\Tto\over\sqrt{1-\Ttz^2}}\nonumber\\
\Uo&=&-{\grad T_1\over a^2\Ttza}.
\label{pertidentify}
\earay
where $\grad$ is an ordinary gradient with respect to the spatial coordinates,
computed as if space is flat. Note that although the perturbations (like the
background) do not obey any
simple equation of state,
nevertheless
$\po/\rhoo\sim 1-\Ttz^2$, which we expect to be very small.

The perturbed Einstein
field equations follow from Eqs. (15.10.29)-(15.10.31) in \cite{Weinberg}:
\baray
-8\pi G(\rhoo-\po)\delta_{ij}&=&
{\nabla^2h_{ij}\over a^2}-{h_{ik,jk}\over a^2}-{h_{jk,ik}\over a^2}+{h_{,ij}\over
a^2}-\ddot h_{ij}-3H\dot h_{ij}-\delta_{ij}H\dot h\nonumber\\
-16\pi G(\rhotz+\ptz){T_{1,i}\over
\Ttza}&=&{\partial(h_{,i}-h_{ki,k})\over\partial t}\nonumber\\
-8\pi G(\rhoo+3\po)&=&\ddot h+2H\dot h,
\label{einpert}
\earay
where $h=\Tr(h_{ij})$. The pertubed equation of motion for the tachyon field
is (from Eq. [\ref{energycons}] or Eq. [15.10.33] in Weinberg)
\be
\dot\rhoo+3H(\rhoo+\po)=-(\rhotz+\ptz)\left({\dot h\over 2}-{T_{1,ii}\over a^2\Ttza}
\right).
\label{energypert}
\ee
Note that we should be able to derive Eq. (\ref{energypert}) from appropriate
manipulation of Eqs. (\ref{einpert}), since the (linearized) Bianchi identities
imply that the (linearized) field equations follow from the divergence of the
(linearized) Einstein equations.

We are interested in scalar modes, so the metric perturbations ought to
reduce to scalar functions; let us assume that (e.g. \cite{BKP})
\be
h_{ij}(\xvec,t)=A(\xvec,t)\delta_{ij}+[B(\xvec,t)]_{,ij}
\,\Rightarrow\,h=3A(\xvec,t)+\nabla^2B(\xvec,t).
\ee
so that Einstein's equations may be rewritten as
\baray
-8\pi G(\rhoo-\po)\delta_{ij}&=&
\left({\nabla^2A\over a^2}-\ddot A-6H\dot A-H\nabla^2\dot B\right)\delta_{ij}
+\left({A\over a^2}-\ddot B-3H\dot B\right)_{,ij}\nonumber\\
-16\pi G(\rhotz+\ptz){T_{1,i}\over\Ttza}&=&(2\dot A)_{,i}\nonumber\\
-8\pi G(\rhoo+3\po)&=& 3\ddot A+6H\dot A+\nabla^2\ddot B+2H\nabla^2\dot B
\nonumber\\
&=&{\partial^2(3A+\nabla^2B)\over\partial t^2}+2H{\partial(3A+\nabla^2B)\over
\partial t}~,
\label{einpert2}
\earay
and we may also rewrite Eq. (\ref{energypert}) as
\be
\dot\rhoo+3H(\rhoo+\po)=-(\rhotz+\ptz)\left[{1\over 2}
\left(3\dot A+\nabla^2\dot B\right)
-{T_{1,ii}\over a^2\Ttza}\right].
\label{energypert2}
\ee
Take the trace of the first of Eqs. (\ref{einpert2}) to find
\baray
-24\pi G(\rhoo-\po)&=&{4\nabla^2A\over a^2}-3\ddot A-18H\dot A-\nabla^2\ddot B
-6H\nabla^2\dot B\nonumber\\
&=&{4\nabla^2A\over a^2}-{\partial^2(3A+\nabla^2B)\over\partial t^2}
-6H{\partial(3A+\nabla^2B)\over\partial t}~,
\label{traceqn}
\earay
then take $\nabla_i\nabla_j$ of the first of Eqs. (\ref{einpert2}) to get
\be
-8\pi G\nabla^2(\rhoo-\po)=\nabla^2\left({2\nabla^2A\over a^2}-\ddot A-6H\dot A
-\nabla^2\ddot B-4H\nabla^2\dot B\right)~,
\ee
and finally subtract ${1\over 3}\nabla^2$ of Eq. (\ref{traceqn}) to find the homogeneous
equation
\be
0=\nabla^4\left({2A\over 3a^2}-{2\ddot B\over 3}-{2H\dot B}\right);
\ee
conclude that (for modes $\propto e^{i\kvec\dotprod\xvec}$ with
$\vert\kvec\vert\neq 0$)
\be
{A\over a^2}=\ddot B+3H\dot B.
\label{asolution}
\ee
Add the last of Eqs. (\ref{einpert2}) to Eq. (\ref{traceqn}), and divide
the result by four:
\be
-8\pi G\rhoo={\nabla^2A\over a^2}-3H\dot A-H\nabla^2\dot B~;
\label{rhoeqn}
\ee
subtract the result from the last of Eqs. (\ref{einpert2}), and
use Eq. (\ref{asolution}) to find
\be
\ddot A+3H\dot A=-8\pi G\po.
\label{peq}
\ee
Following Ref. \cite{BKP}, we can shall use the last three equations, supplemented
by the middle of Eqs. (\ref{einpert2}), to derive gauge invariant equations
describing the evolution of cosmological density perturbations.

Consider the gauge invariant combination (for proof, see \cite{BKP})
\be
\phih={1\over 2}\left(A-a^2H\dot B\right)={1\over 2}a^2(\ddot B+2H\dot B)
\equiv {1\over 2}\dot\psih,
\ee
where $\psih=a^2\dot B$. We can rewrite Eq. (\ref{peq}) as
\be
\stackrel{\cdots}{\psi}_H+4H\ddot\psih+(2\dot H+3H^2)\dot\psih
+(\ddot H+3H\dot H)\psih=-8\pi G\po,
\ee
or, in terms of $\phih$,
\be
2\ddot\phih+8H\dot\phih+2(2\dot H+3H^2)\phih+
(\ddot H+3H\dot H)\psih=-8\pi G\po.
\ee
To eliminate $\psih$ from this equation, use the middle of Eqs. (\ref{einpert2}),
which can be integrated once spatially to yield
\be
\dot A=-8\pi G(\rhotz+\ptz){T_1\over\Ttza};
\ee
Eq. (\ref{asolution}) implies that $A=\dot\psih+H\psih$, and therefore
\be
\ddot\psih+H\dot\psih+\dot H\psih=-8\pi G(\rhotz+\ptz){T_1\over\Ttza},
\ee
which we can solve for $\psih$:
\baray
\psih&=&-{8\pi G(\rhotz+\ptz)T_1\over\dot H\Ttza}
-{H\dot\psih\over\dot H}-{\ddot\psih\over\dot H}
\nonumber\\
&=&-{8\pi G(\rhotz+\ptz)T_1\over\dot H\Ttza}
-{2H\phih\over\dot H}-{2\dot\phih\over\dot H}~.
\earay
Substituting for $\psih$ results in the equation
\be
\ddot\phih+\left(H-{\ddot H\over \dot H}\right)\dot\phih
+\left(2\dot H-{H\ddot H\over \dot H}\right)\phih
=-4\pi G\left[\po-{(\rhotz+\ptz)(\ddot H +3H\dot H)T_1\over
\dot H\Ttza}\right].
\label{phiheqn}
\ee
We can get a second equation for $\phih$ from Eqs. (\ref{rhoeqn}),
which may be rewritten in the form
\be
{\nabla^2\phih\over a^2}=-4\pi G\rhoo+{3H\dot A\over 2}
=-4\pi G\left[\rhoo+{3H(\rhotz+\ptz)T_1\over\Ttza}\right]~,
\label{phiheqn2}
\ee
where we have used Eq. (\ref{asolution}) to substitute for $\dot A$.
\footnote{The source term in Eq. (\ref{phiheqn2}) is easily shown
to be the gauge invariant density perturbation.}

To go further, we need to return to the equations for the cosmological
background and compute various derivatives of the expansion rate, $H$,
starting from Eqs. (\ref{bckgnd}). We then find
\baray
\dot H&=&-4\pi G(\rhotz+\ptz)\nonumber\\
\ddot H&=&12\pi GH(\rhotz+\ptz)(1+c_s^2)\nonumber\\
{\ddot H+3H\dot H\over\dot H}&=&-3Hc_s^2,
\label{Hderivs}
\earay
where $c_s^2\equiv\dot\ptz/\dot\rhotz$. For rolling tachyons we have
\baray
\dot\ptz&=&-\dot\rhotz(1-\Ttz^2)+2\rhotz\Ttz\ddot T_0
=(1-\Ttz^2)\left[\dot\rhotz-{2\rhotz V^\prime(T_0)\dot T_0\over V(T_0)}
\right]\nonumber\\
c_s^2&=&(1-\Ttz^2)\left[1-{2\rhotz V^\prime(T_0)\dot T_0\over V(T_0)\dot\rhotz}\right]
=(1-\Ttz^2)\left[1+{2V^\prime(T_0) \over 3\Ttz V(T_0)H}\right];
\label{cseqn}
\earay
although the second factor in the above can be large, we still expect very small
$c_s^2$ (e.g. $1-\Ttz^2$ is exponentially small for
$V(T) = V_0 \exp(-T/\tau)$
at
late times).\footnote{With these results it is easy to show that the source term
in Eq. (\ref{phiheqn}) is just the gauge invariant pressure perturbation.}
We can combine Eqs. (\ref{phiheqn}) and (\ref{phiheqn2}) into a single
differential equation for $\phih$ as follows. First, multiply Eq. (\ref{phiheqn2})
by $1-\Ttz^2$ to find
\baray
(1-\Ttz^2){\nabla^2\phih\over a^2}&=&
=-4\pi G(1-\Ttz^2)\left[\rhoo-\dot\rhotz{T_1\over\Ttza}\right]
\nonumber\\
&=&-4\pi G\left\{\po+\left[2V^\prime(T_0)\Ttza\sqrt{1-\Ttz^2}
-\dot\rhotz(1-\Ttz^2)\right]{T_1\over\Ttza}\right\}\nonumber\\
&=&-4\pi G\left(\po-{c_s^2\dot\rhotz T_1\over\Ttza}\right)
\nonumber\\
&=&-4\pi G\left(\po-{\dot\ptz T_1\over\Ttza}\right),
\earay
where, to get the final version of this equation, we used Eq. (\ref{pertidentify})
and Eq. (\ref{cseqn}). But Eqs. (\ref{Hderivs}) can be used to rewrite
Eq. (\ref{phiheqn}) as
\be
\ddot\phih+\left(H-{\ddot H\over \dot H}\right)\dot\phih
+\left(2\dot H-{H\ddot H\over \dot H}\right)\phih
=-4\pi G\left(\po-{\dot\ptz T_1\over
\Ttza}\right);
\ee
combining the last two results we find the homogeneous equation
\be
\ddot\phih+\left(H-{\ddot H\over \dot H}\right)\dot\phih
+\left(2\dot H-{H\ddot H\over \dot H}\right)\phih
-(1-\Ttz^2){\nabla^2\phih\over a^2}=0.
\label{phiheqn3}
\ee
The last term is very small because $1-\Ttz^2\ll 1$, so that the
effective Jeans length for density perturbations is only of order
$L_{J,t}=H^{-1}\sqrt{1-\Ttz^2}$. Thus, tachyon matter {\it can}
cluster on all but very small length scales, once $\Ttz\to 1$.
At earlier times, when $\Ttz<1$, clustering is prevented on
scales smaller than $\sim H^{-1}$; if the tachyon field rolls
slowly, the growth of subhorizon tachyon matter perturbations
is suppressed, just as for quintessence. 

For perturbations with
proper sizes larger than $L_{J,t}$, Eq. (\ref{phiheqn3}) implies
that
\be
\phih=C_0+{C_1\over t^{5/3}}
\ee
during epochs when the tachyon density dominates and decreases
$\sim a^{-3}$ to a very good approximation. Then from Eq. (\ref{phiheqn2}),
the gauge-invariant tachyon matter density contrast
\be
\Delta^{(t)}\equiv{\rhoo-\dot\rhotz T_1/\Ttza\over\rhotz}
\propto{\phih\over a^2\rhotz}=C_0t^{2/3}+{C_1\over t},
\ee
which is a linear combination of the well-known growing and shrinking
modes for perturbations of nonrelativistic matter. (See e.g. Weinberg
\cite{Weinberg}, \S~15.9, 15.10, or Peebles \cite{Peebles}, \S~11.)

\section{Conclusions}

We have shown that tachyon matter is an acceptable candidate
for the dark matter of the universe, provided the initial conditions
for the rolling tachyon field are fine tuned. In particular, the mass
density in tachyonic form decreases $\propto a^{-3}$ at late times,
as has been observed by others \cite{{gibbons},{FKS}}, but only
has an acceptable value today provided that the initial value of
the tachyon field is in a rather small, and special range (e.g. \cite{sen3} and
\S~\ref{bcosmo}), assuming that standard big bang cosmology proceeds
uninterrupted after the tachyon begins to roll.
If the tachyon field begins rolling at too small a
value, then the density today would turn out to be unacceptably
large, given the temperature of the universe today -- a well-measured
cosmological parameter -- and a large range of reasonable values
for the temperature of the universe when the tachyon starts to roll. 
If the tachyon field begins rolling at too large a value, then its
density today is inconsequential (i.e. $\Omega_{T,0}\ll 1$), so it
plays a relatively harmless, but uninteresting role in cosmology.
Since the required starting value of $\Tz$ estimated in \S~\ref{bcosmo}
is rather large (about $60\tau$ if the fundamental scale is the Planck
mass, or about $23\tau$ if it is TeV, for an exponential potential
$V(T)\propto \exp(-T/\tau)$ and $\Omega_{T,0}\sim 1$), small fractional
differences have substantial consequences for the subsequent evolution
of the universe. (This fine-tuning avoids the over-growth of density fluctuations
noted in Ref. \cite{FKS}, at the price of possibly unnatural initial
conditions.)
Why the tachyon field begins rolling from a small
range of initial values far from unity must be explained if the rolling
tachyon is to be regarded as a natural candidate for the dark
matter of the universe.

Alternatively, the tachyon field could have started rolling from
``generic'' initial conditions provided that a second period of
inflation occurred after the tachyon field had rolled for awhile.
In that case, there is still a fine tuning issue, since the value
of $\Temp_{start}a_{start}^{3/4}$ must be adjusted precisely
(to a value of about $0.003\Omega_{T,0}^{1/4}$ eV; see \S~\ref
{bcosmo}) in order for $\Omega_{T,0}\sim 1$. For a given value
of $\Temp_{start}$ (e.g. the Planck scale, or whatever the fundamental
mass scale of string theory might be), this implies a very 
small range of values for the expansion factor during the second
inflationary phase. For a smaller expansion factor, the density
of tachyons could not satisfy $\Omega_{T,0}<1$, whereas for a larger
factor, the tachyon density today would be too low to be of
any importance to cosmology.

\acknowledgments

We would like to thank Lam Hui, Bhuvnesh Jain,
Ashoke Sen, and Max Tegmark for discussions.
The work of GS was supported in part by the DOE grants
DE-FG02-95ER40893, DE-EY-76-02-3071 and the University of Pennsylvania
School of Arts and Sciences Dean's funds. G.S. also
thanks the Michigan Center for Theoretical Physics for hospitality
while this paper was written.

\end{document}